\documentclass[conference]{IEEEtran}
\IEEEoverridecommandlockouts
\usepackage{marvosym}

\usepackage{xcolor}
\usepackage[normalem]{ulem}
\usepackage{subfigure}
\usepackage{minted}
\usepackage{tikz}
\usepackage{amsmath,amssymb}
\usepackage{graphicx}
\usepackage{booktabs}
\usepackage{hyperref}
\usepackage{cleveref}
\usepackage{microtype}
\usepackage{enumitem}
\usepackage{multirow}
\usepackage{float}
\usepackage{footnote}
\usepackage[compress]{cite}
\usepackage{graphicx}
\usepackage{blindtext}
\usepackage{hyperref}
\usepackage[utf8]{inputenc}
\usepackage{amssymb}
\usepackage{float}
\usepackage{enumitem}
\usepackage{algorithm}
\usepackage{algpseudocode}
\usepackage{amsmath}
\usepackage{makecell}
\usepackage{footnote}
\usepackage{amsmath}
\usepackage{listings}
\usepackage{color}
\usepackage{longtable}
\usepackage{caption} 
\usepackage{amssymb}
\usepackage{xcolor}
\definecolor{verylightgray}{gray}{0.95}

\newtheorem{remark}{Remark}

\newcommand{\tNAME}{\textsc{PrivMark}}

\newcommand{\todo}[1]{%
    \mbox{}
    \marginpar{%
        \colorbox{yellow!100}{\textcolor{white}{TODO}}%
        \vspace*{-22pt}
    }%
    \textcolor{red}{\;\;\;\;#1}%
}

\definecolor{eclipseStrings}{RGB}{42,0.0,255}
\definecolor{eclipseKeywords}{RGB}{127,0,85}
\colorlet{numb}{magenta!60!black}

\definecolor{eclipseStrings}{RGB}{42,0.0,255}
\definecolor{eclipseKeywords}{RGB}{127,0,85}
\colorlet{numb}{magenta!60!black}

\hypersetup{
  colorlinks   = true,    
  urlcolor     = blue,    
  linkcolor    = blue,    
  citecolor    = blue      
}

\def\BibTeX{{\rm B\kern-.05em{\sc i\kern-.025em b}\kern-.08em
    T\kern-.1667em\lower.7ex\hbox{E}\kern-.125emX}}
\begin{document}

\title{\tNAME{}: Private Large Language Models Watermarking with MPC}

\author{
Thomas Fargues\textsuperscript{†},
Ye Dong\textsuperscript{‡,(\Letter)},
Tianwei Zhang\textsuperscript{§},
Jin Song Dong\textsuperscript{‡} \\
\textsuperscript{†}Télécom SudParis, France, \textsuperscript{‡}National University of Singapore, \textsuperscript{§}Nanyang Technological University, Singapore \\
{{thomas.fargues@telecom-sudparis.eu}, {dongye@nus.edu.sg}, {tianwei.zhang@ntu.edu.sg}, {csdjs@nus.edu.sg}}
}

\maketitle

\begin{abstract}
The rapid growth of Large Language Models (LLMs) has highlighted the pressing need for reliable mechanisms to verify content ownership and ensure traceability. Watermarking offers a promising path forward, but it remains limited by privacy concerns in sensitive scenarios, as traditional approaches often require direct access to a model’s parameters or its training data. 
In this work, we propose a secure multi-party computation (MPC)-based private LLMs watermarking framework, \tNAME{}, to address the concerns.
Concretely, we investigate PostMark (EMNLP'2024), one of the state-of-the-art LLMs Watermarking methods, and formulate its basic operations.
Then, we construct efficient protocols for these operations using the MPC primitives in a black-box manner.
In this way, \tNAME{} enables multiple parties to collaboratively watermark an LLM’s output without exposing the model’s weights to any single computing party. 
We implement \tNAME{} using SecretFlow-SPU (USENIX ATC'2023) and evaluate its performance using the ABY3 (CCS'2018) backend. The experimental results show that \tNAME{} achieves semantically identical results compared to the plaintext baseline without MPC and is resistant against paraphrasing and removing attacks with reasonable efficiency.
\end{abstract}


\section{Introduction}\label{sec:introduction}
Large Language Models (LLMs) such as ChatGPT~\cite{chatgpt} have demonstrated remarkable capabilities in a wide range of applications from content creation to sophisticated conversational agents. As these models become more powerful and integrated into commercial and critical systems, questions of authorship, intellectual property, and misuse become increasingly salient~\cite{kendall2024risks}. Malicious actors can exploit these models to generate misinformation, plagiarized content, or other harmful text, making it difficult to trace the origin of the generated output. Consequently, developing reliable techniques to identify machine-generated text and verify model ownership is a critical area of research.
Digital watermarking~\cite{pan2024markllm,liang2024watermarking} offers a compelling solution to these challenges. By subtly embedding a secret signal, a ``watermark", into the generated text, model owners can later prove their ownership or identify the source of a particular piece of content. Watermarking techniques for LLMs can be broadly categorized into two groups: those that modify the training process or model architecture, for instance, by fine-tuning the model to embed a secret signature within its parameters, effectively teaching it a ``dialect" that is statistically detectable \cite{abdelnabi2021adversarialwatermarkingtransformertracing}, and those that are applied during or after the text generation phase. The latter is a more common approach where the watermark is applied during inference; for example, a secret key can be used to partition the model's vocabulary into ``green" and ``red" lists, and the generation algorithm is then steered to preferentially select words from the green list, creating a detectable statistical bias in the output text \cite{kirchenbauer2024watermarklargelanguagemodels, zhao2023provable}. Post-generation watermarking schemes, such as the PostMark~\cite{chang2024postmarkrobustblackboxwatermark}, are particularly attractive as they do not require costly retraining and can be applied to pre-trained LLMs directly.

However, a significant barrier to the widespread adoption of LLM watermarking is the issue of privacy. In many real-world scenarios, the LLM itself, the data it processes, and the watermark key are all sensitive assets. For instance, a consortium of companies might want to co-own and use an LLM for internal data analysis without exposing their proprietary model or confidential data to each other. In such collaborative or privacy-sensitive settings, traditional watermarking methods that require a central party to have full access to all components are not viable.

\noindent{\bf Our Contributions.}
To bridge this gap, we propose the first, to our best knowledge, private LLM watermarking framework, \tNAME{}, by integrating Secure Multiparty Computation (MPC)~\cite{yao1986generate,goldreich1987} primitives into LLMs watermarking progress to protect privacy.  
MPC is a cryptographic technology that enables multiple parties to jointly compute a function over their private inputs without revealing those inputs to one another.
Concretely, we secret-share the private LLM weights, watermarking parameters, and user data among different non-colluding computing parties.
By making use of advanced three-party (3PC) primitives in the Arithmetic Black Box (ABB) model~\cite{canetti2001universally}, we construct efficient 3PC protocols for both inserting and detecting watermarks. 
Our \tNAME{} represents a significant step towards enabling secure and accountable LLMs.
In summary, our contributions are threefold: 
\begin{itemize}[leftmargin=10pt]
\item We conduct a systematic investigation and comparison of existing LLM watermarking approaches. For generality, we employ PostMark~\cite{chang2024postmarkrobustblackboxwatermark}, one state-of-the-art technique, as the representative watermarking method.
\item We analyze the workflow of PostMark, decomposing it into fundamental operations. We then develop efficient and secure protocols for these operations by leveraging 3PC primitives within the ABB model.
\item We implement the proposed protocol in SecretFlow-SPU using the ABY3 protocol~\cite{aby3}. Our evaluation demonstrates its practical feasibility and shows that it effectively preserves privacy while maintaining both watermark integrity and LLM utility. Concretely, \tNAME{} achieves comparable robustness against paraphrasing and removing attacks \S\ref{subsec:robustness} with reasonable overhead.  
\end{itemize}


\noindent{\bf Organization.}
In \S\ref{sec:relatedwork}, we summarize related works. Then, we introduce PostMark and 3PC in \S\ref{sec:background}.
The concrete design of \tNAME{} is shown in \S\ref{sec:design}.
Finally, we show experimental evaluations in \S\ref{sec:experiment} and conclude this work in \S\ref{sec:conclusion}.

\section{Related Work}\label{sec:relatedwork}

We summarize recent watermarking approaches and MPC-based private inference frameworks for LLMs.

\subsection{Watermarking of Large Language Models}
The proliferation of powerful LLMs has spurred research into techniques for identifying machine-generated text, which can be applied during training or post-generation \cite{liang2024watermarkingtechniqueslargelanguage}. Training-time methods embed a signal by modifying the model's architecture or fine-tuning process, but this is often computationally expensive and impractical for proprietary, pre-trained models.
Post-generation (or black-box) watermarking methods are more flexible, as they operate on the model's output probabilities without requiring access to its internal weights. Also, this method is more efficient~\cite{liang2024watermarkingtechniqueslargelanguage}.
As shown in Table~\ref{tab:llm_watermarking_part1}, we give a holistic comparison of representative methods in terms of fidelity, robustness, efficiency, and undetectability, which are the most important properties of Watermarking methods.
\begin{table}[ht]
\centering
\caption{Comparison of LLM watermarking Methods.}
\vspace{-5pt}
\label{tab:llm_watermarking_part1}
\footnotesize
\resizebox{\linewidth}{!}{
\begin{tabular}{|l|c|c|c|c|}
\hline
\textbf{Method} & \textbf{Fidelity} & \textbf{Robustness} & \textbf{Efficiency} & \textbf{Undetectability} \\
\hline

\textbf{Unicode \cite{sato2023embarrassinglysimpletextwatermarks}} &
  \begin{tabular}[c]{@{}c@{}}$\bigstar\bigstar\bigstar$ \\ Preserves semantics\end{tabular} &
  \begin{tabular}[c]{@{}c@{}}$\bigstar$ \\ Very Weak\end{tabular} &
  \begin{tabular}[c]{@{}c@{}}$\bigstar\bigstar\bigstar$ \\ Extremely High\end{tabular} &
   \begin{tabular}[c]{@{}c@{}}$\bigstar\bigstar$ \\ Machine-detectable\end{tabular} \\ \hline

\textbf{Yang et al. \cite{yang2023watermarkingtextgeneratedblackbox}} &
  \begin{tabular}[c]{@{}c@{}}$\bigstar\bigstar\bigstar$ \\ High Fidelity\end{tabular} &
  \begin{tabular}[c]{@{}c@{}}$\bigstar\bigstar$ \\ Moderate-to-High\end{tabular} &
  \begin{tabular}[c]{@{}c@{}}$\bigstar\bigstar$ \\ Moderate\end{tabular} & 
  \begin{tabular}[c]{@{}c@{}}$\bigstar\bigstar\bigstar$ \\ Imperceptible\end{tabular}\\ \hline

\textbf{DeepTextMark \cite{munyer2024deeptextmarkdeeplearningdriventext}} &
  \begin{tabular}[c]{@{}c@{}}$\bigstar\bigstar\bigstar$ \\ High Fidelity\end{tabular} &
  \begin{tabular}[c]{@{}c@{}}$\bigstar\bigstar\bigstar$ \\ High\end{tabular} &
  \begin{tabular}[c]{@{}c@{}}$\bigstar\bigstar$ \\ Good\end{tabular} &
  \begin{tabular}[c]{@{}c@{}}$\bigstar\bigstar\bigstar$ \\ Imperceptible\end{tabular} \\ \hline

\textbf{PostMark \cite{chang2024postmarkrobustblackboxwatermark}} &
  \begin{tabular}[c]{@{}c@{}}$\bigstar\bigstar$ \\ Medium\end{tabular} &
  \begin{tabular}[c]{@{}c@{}}$\bigstar\bigstar\bigstar\bigstar$ \\ Very High\end{tabular} &
  \begin{tabular}[c]{@{}c@{}}$\bigstar$ \\ Lower\end{tabular} &
  \begin{tabular}[c]{@{}c@{}}$\bigstar\bigstar\bigstar$ \\ Imperceptible\end{tabular} \\ \hline

\textbf{REMARK-LLM \cite{299615}} &
  \begin{tabular}[c]{@{}c@{}}$\bigstar\bigstar\bigstar$ \\ High Fidelity\end{tabular} &
  \begin{tabular}[c]{@{}c@{}}$\bigstar\bigstar\bigstar$ \\ High\end{tabular} &
  \begin{tabular}[c]{@{}c@{}}$\bigstar\bigstar$ \\ Moderate\end{tabular} &
  \begin{tabular}[c]{@{}c@{}}$\bigstar\bigstar\bigstar$ \\ Imperceptible\end{tabular}\\ \hline

\end{tabular}}
\vspace{-10pt}
\end{table}

From our analysis, three methods emerged as strong candidates with high resilience to paraphrasing attacks: DeepTextMark \cite{munyer2024deeptextmarkdeeplearningdriventext}, PostMark \cite{chang2024postmarkrobustblackboxwatermark}, and REMARK-LLM \cite{299615}. However, the PostMark algorithm \cite{chang2024postmarkrobustblackboxwatermark} is a prominent example of a robust black-box technique. It works by using a secret table to pseudo-randomly select words to insert into the text of the embedded watermark.
This makes PostMark an ideal candidate for pioneering the integration of watermarking into a privacy-preserving scenario.



\subsection{MPC-based Private LLMs Inference}
Secure Multi-Party Computation (MPC) provides a cryptographic foundation for parties to jointly compute a function on their private data~\cite{yao1986generate, goldreich1987}. 
Recently, the focus has shifted towards the unique challenges posed by large-scale Transformer architectures. Several works have proposed MPC-based solutions for secure Transformer inference \cite{li2023mpcformerfastperformantprivate, hao2022iron,dong2023pumasecureinferencellama7b,lu2023bumblebee,pang2023bolt,gupta2023sigma} in two or three-party. These advancements have concentrated almost exclusively on the core task of secure inference.

\section{Background}\label{sec:background}
In this section, we introduce PostMark WaterMarking and replicated secret sharing-based three-party computation (3PC).

\subsection{Revisiting PostMark}\label{sec:postmark_solution}

PostMark \cite{chang2024postmarkrobustblackboxwatermark} is a post-generation watermarking algorithm designed for black-box LLMs. The algorithm consists of two main procedures: insertion and detection.

\begin{figure}
    \centering
    \includegraphics[width=\linewidth]{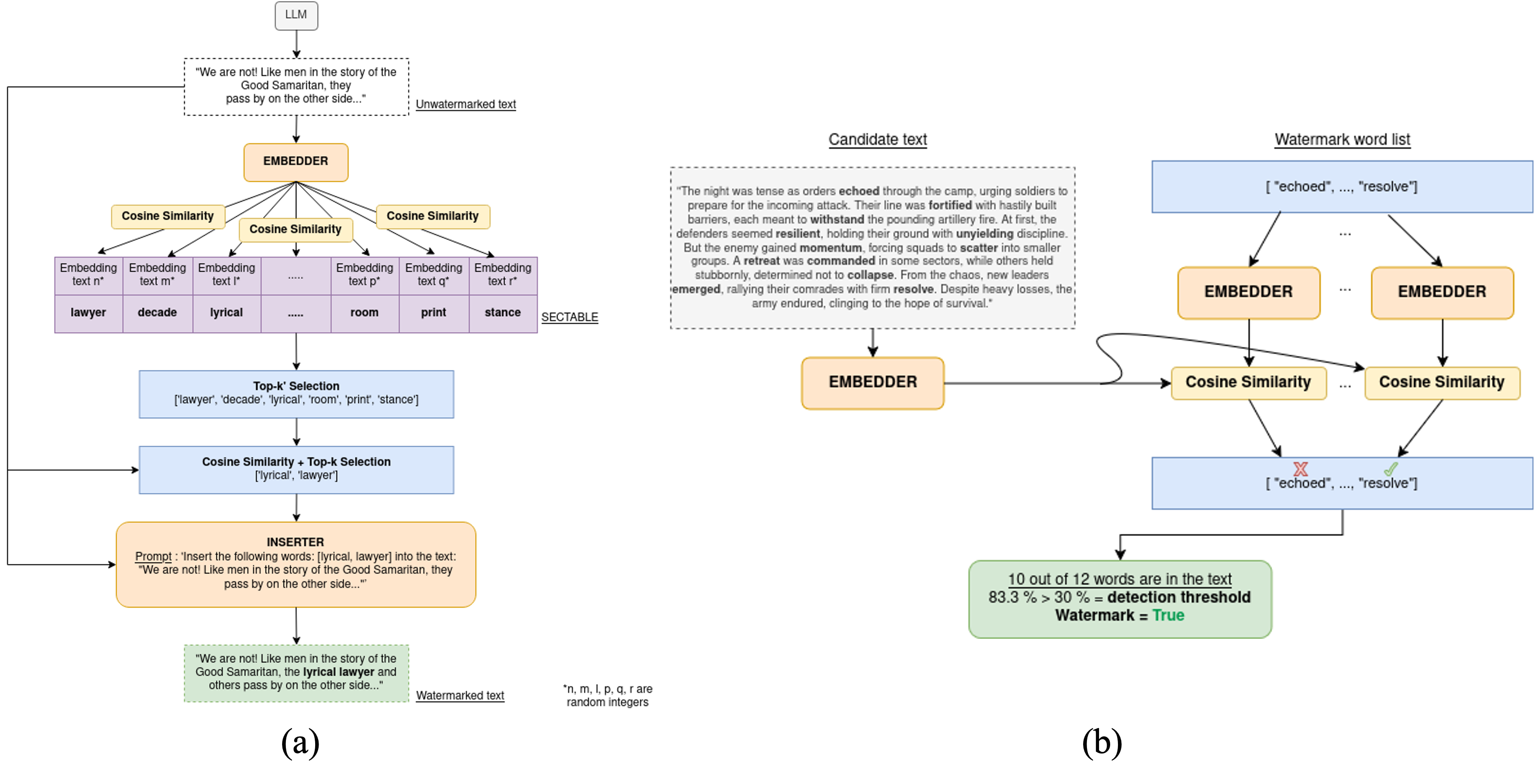}  
    \vspace{-20pt}
    \caption{PostMark watermarking insertion (a) and detection (b).}
    \label{fig:postmark_insertion}
    \vspace{-10pt}
\end{figure}

\noindent{\bf Insertion.} As shown in Figure \ref{fig:postmark_insertion} (a), PostMark embeds watermarks with \texttt{EMBEDDER} by leveraging semantic similarity between the input text and a set of carefully selected watermark words. The process begins by generating a semantic embedding of the original text using a pretrained embedding model (such as OpenAI’s text-embedding-3-large). This embedding is compared to a secret word embedding table, \texttt{SECTABLE}\footnote{This table is built from external files: a vocabulary $V$ and a set of embeddings $D$, we detail this in \S\ref{sec:watermark_insert}}, which contains embeddings randomly assigned to a curated vocabulary of content words. And after, the top-k similar words from \texttt{SECTABLE} are selected to serve as watermark candidates.

An instruction-following language model (\texttt{INSERTER}), such as GPT-4o, is then prompted to rewrite the original text while incorporating the selected watermark words in a natural and fluent way, preserving the original semantics. 

\noindent{\bf Detection.} For detection (Figure \ref{fig:postmark_insertion} (b)), the system re-embeds the suspect text, reconstructs the list of candidate watermark words based on similarity to a candidate list, and checks how many of those words appear in the text. If the overlap exceeds a certain threshold, the text is considered watermarked.

PostMark is designed to be robust against paraphrasing attacks, since it conditions the watermark on the text's semantic embedding, which remains relatively stable under paraphrasing.

\subsection{Secure Multiparty Computation \& Threat Model}

\subsubsection{MPC Preliminaries}

\tNAME{} is built upon \textbf{replicated secret sharing-based three-party computation (3PC)}~\cite{mohassel2018aby3}. A secret value $x$ is split into three shares, $(x_1, x_2, x_3)$, such that $x = x_1 + x_2 + x_3$ in ring $\mathbb{Z}_{2^\ell}$. These shares are then distributed among three computing parties as: $P_1$ holds $(x_1, x_2)$, $P_2$ holds $(x_2, x_3)$, and $P_3$ holds $(x_3, x_1)$.

This structure allows parties to perform computations on the shares directly. \textbf{Linear operations}, such as addition, can be performed locally by each party on their respective shares without any communication. \textbf{Non-linear operations}, such as multiplication, require dedicated and efficient interactive protocols where parties exchange messages to compute shares of the product.
Existing works~\cite{spu,aby3,dong2023pumasecureinferencellama7b} have proposed efficient protocols for widely used arithmetic and boolean functions (including comparison, selection, etc) and even secure inference of neural networks. We use them in the ABB model to construct \tNAME{}.

\subsubsection{Threat Model}
We follow the standard \textbf{honest-but-curious} security and build \tNAME{} upon ABY3 protocol. Under this model, all computing parties are expected to honestly follow the protocol's instructions, but they may attempt to learn additional information about other parties' private data from the messages they receive during the computation. Our security guarantees hold as long as at most one of the three parties is corrupted by the passive adversary.


\section{Design of \tNAME{}}\label{sec:design}

\subsection{Overview of Our Design}\label{sec:overview}
\begin{figure}[!t]
    \centering
    \subfigure[]{
    \includegraphics[width=1\linewidth]{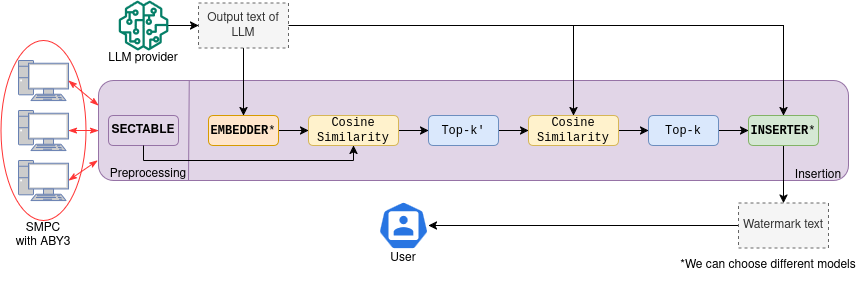}
    }
    \vspace{-5pt}
    \subfigure[]{
    \includegraphics[width=1\linewidth]{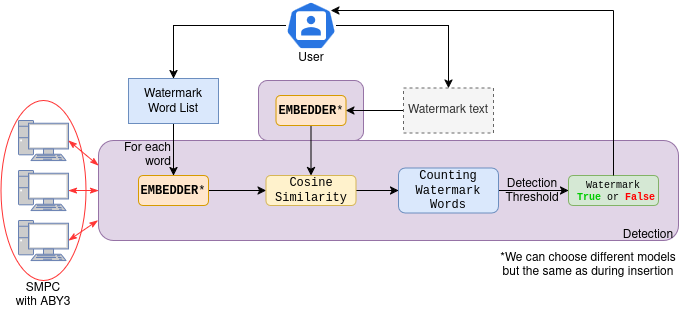}
    }
    \vspace{-5pt}
    \caption{Illustration of PrivMark insertion (a) and detection (b).}
    \label{fig:privmark_insertion}
    \vspace{-10pt}
\end{figure}

To achieve secure watermarking, \tNAME{} uses three non-colluding computing parties to execute ABY3~\cite{mohassel2018aby3} to insert and detect the watermarking. \tNAME{} is designed to follow the same workflow as PostMark, but with a focus on privacy preservation. The workflow is illustrated in Figure \ref{fig:privmark_insertion}:
\begin{itemize}[leftmargin=10pt]
    \item For secure evaluations of \texttt{EMBEDDER} and \texttt{INSERTER}, we utilize ABY3-based secure inference solution~\cite{dong2023pumasecureinferencellama7b} to achieve secure embedding and LLM-based rewriting.

    \item For other operations of PostMark, we invoke the 3PC primitives provided by SecretFlow-SPU~\cite{spu}.
\end{itemize}

The solution can be understood as a workflow involving the client, the LLM provider, and three computing parties that jointly process the watermarking tasks.  

\smallskip
\noindent\textbf{Scenario.}  
Consider a company that develops innovative product designs and relies on an external LLM provider to assist in drafting technical reports and marketing documents. These documents are part of the company’s intellectual property (IP) and must be protected against unauthorized redistribution or misuse. With \tNAME{}, the company can ensure that all LLM outputs contain invisible watermarks. Even if competitors or external parties obtain the generated documents, the watermark allows the company to prove ownership and trace the origin of the content. Importantly, with MPC, the watermarking process does not expose the company’s sensitive data nor the parameters of the LLMs and watermarking algorithms.  

\smallskip
\noindent\textbf{Insertion Workflow.}  
The watermark insertion proceeds as follows:  \textbf{1) Client request}: The client sends a query to the LLM provider.  
\textbf{2) LLM output}: The LLM provider generates an answer \(t\), which serves as input to \tNAME{}.  
\textbf{3) Initialization}: Three computing parties start the secure computation and jointly generate the \texttt{SECTABLE}.  
\textbf{4) Secure embedding}: Using the \texttt{EMBEDDER}, the three parties securely embed \(t\) as $\langle e_t\rangle$.  
\textbf{5) First similarity search}: The parties compute cosine similarity between $\langle e_t\rangle$ and \texttt{SECTABLE}, then select the top-$k'$ candidate words.  
\textbf{6) Watermark selection}: Watermark words are securely selected. Using \texttt{INSERTER}, they are injected into \(t\) via prompt modification.  
\textbf{7) Delivery}: The final watermarked text is returned to the user.  
Figure \ref{fig:privmark_insertion} (a) and Algorithm \ref{alg:insertion} show details.  

\smallskip
\noindent\textbf{Detection Workflow.}  
The watermark verification process follows these steps:  
\textbf{1) Client submission}: The user provides \tNAME{} with a candidate text and the list of watermark words.
\textbf{2) Secure embedding}: The three computing parties embed both the candidate text and watermark word list using the \texttt{EMBEDDER}.  
\textbf{3) Similarity computation}: For each word, the three parties compute the cosine similarity securely.  
\textbf{4) Watermark identification}: The system counts how many words match the watermark pattern.  
\textbf{5) Threshold decision}: Based on a detection threshold, \tNAME{} determines whether the candidate text is watermarked.  
\textbf{6) Result delivery}: The decision (watermarked / not watermarked) is returned to the user.  
All details are shown in Figure \ref{fig:privmark_insertion} (b) and Algorithm \ref{alg:detection}.


\subsection{Watermark Insertion}\label{sec:watermark_insert}
\noindent\textbf{Embedder.} The \texttt{EMBEDDER} accounts for converting both individual words and entire documents into a high-dimensional numerical representation. In \tNAME{}, we employ the open-sourced \textit{intfloat/e5-base}~\cite{wang2022text} for the compilation with SPU.

\noindent \textbf{SecTable}: PostMark's central concept is to have an LLM subtly embed a pre-selected list of watermark words into text. This is done without significantly altering the text's quality or meaning. The watermark words themselves are chosen by comparing the text's embedding with a \texttt{SECTABLE}, which is a word embedding table, using cosine similarity. \texttt{SECTABLE}'s creation involves two key steps, explained further below.

\begin{itemize}[leftmargin=10pt]
    \item \underline{Step 1: Vocabulary Selection \(V\)}

The initial step is to establish a suitable vocabulary, \(V\). This process begins with the WikiText-103 corpus \cite{merity2016pointersentinelmixturemodels} as a foundational vocabulary source. To ensure that the watermarking process does not introduce nonsensical or out-of-place words, a filtering procedure is applied. This refinement involves the removal of: function words, proper nouns, infrequent and rare words. For our experiments \S\ref{sec:experiment}, we used \(V\) provided by PostMark \cite{chang2024postmarkrobustblackboxwatermark}.
The resultant curated set of words constitutes vocabulary \(V\).

\item \underline{Step 2: Mapping Vocabulary to Embeddings}

The second step focuses on mapping the words in \(V\) to their corresponding embeddings in a secure manner. To prevent adversarial recovery of the embedding table, \texttt{SECTABLE} is constructed by randomly assigning each word in the vocabulary to an embedding during a prepossessing step. This resulting map effectively serves as a cryptographic key.

The procedure is as follows:
\begin{enumerate}
    \item A set of embeddings, \(D\), is generated by applying an \texttt{EMBEDDER} to a collection of random documents. For our experiments \S\ref{sec:experiment}, we used a \(D\) created with \texttt{wikitext-103-raw-v1} \cite{merity2016pointersentinelmixturemodels} and \textit{intfloat/e5-base}\cite{wang2022text}, but in practice, these can be random 250-word excerpts from a large dataset like the RedPajama dataset as shown in PostMark \cite{chang2024postmarkrobustblackboxwatermark}.
    
    \item Each word in the vocabulary \(V\) is then randomly mapped to a unique document embedding from the set \(D\). This mapping constitutes the final \texttt{SECTABLE}.
    
\end{enumerate}
\end{itemize}

\begin{remark}
For security, we partition data across the three computing parties: one party stores the embedding set \(D\) and another stores the vocabulary \(V\). The parties then jointly and securely compute \texttt{SECTABLE} using ABY3 so that neither \(D\) nor \(V\) is revealed in the clear. Secure construction of \texttt{SECTABLE} is essential because the table constitutes the scheme's trust anchor and effectively serves as the cryptographic key for the watermarking mechanism; any leakage during its generation would compromise the protection.

It is also important to mention that the party who holds the vocabulary \(V\) must be trustworthy because, it stores the watermarking words to insert. A malicious party may generate a vocabulary \(V\) that is unusable for the watermarking solution.
 \end{remark}

\begin{remark}
 Also, a more straightforward approach might be to use the \texttt{EMBEDDER}'s inherent word embeddings directly. However, this method is eschewed because it is vulnerable to attacks and has been shown to be less effective. The reduced effectiveness is partly due to the high probability of many potential watermark words already existing in the input text prior to the watermarking process.
 \end{remark}

\noindent {\bf Words Selection.} Then, we compare $\langle\text{SECTABLE}\rangle$ and the embedding text $\langle e_t\rangle$ with the function \textit{cosine similarity} \ref{alg:insertion} (l.9). It is a metric two to quantify the semantic similarity between the texts embeddings in this example. The resulting score ranges from -1 (semantically opposite) to 1 (semantically identical), with scores closer to 1 indicating a higher degree of similarity. The formula of \textit{cosine similarity} is : 
\begin{equation} 
\text{CosSim(A,B)} = \frac{A \cdot B}{\|A\| \|B\|} = \frac{\sum_{i=1}^{n} A_{i} B_{i}}{\sqrt{\sum_{i=1}^{n} A_{i}^2} \sqrt{\sum_{i=1}^{n} B_{i}^2}}
\end{equation}
Afterwards, we need to select the \textit{topk} with \(k'\) most similar words, where we choose the $k'$ results with the highest score based on the \texttt{SECTABLE}, we pick the top-$k'$ words.
Next, we embed the top k' words and compute their cosine similarity with the original text embedding $\langle e_t\rangle$.
Finally, we select the \texttt{Top-$k$} with \(k\) results to get our watermark words to insert.

\noindent {\bf Inserter}: The \texttt{INSERTER}'s purpose is to link words from the watermark list into the original text by rewriting it, through another LLM. This can be completed by an existing secure inference solution like PUMA~\cite{dong2023pumasecureinferencellama7b}. 
Putting it all together, we give protocol $\Pi_{{\rm MarkIn}}$ for watermark insertion in algorithm~\ref{alg:insertion}.

\begin{savenotes}
\begin{algorithm}[H]
\caption{Protocol $\Pi_{{\rm MarkIn}}$ for Watermark Insertion}
\label{alg:insertion}
\begin{algorithmic}[1]
\Require Party \(P_1\) holds plaintext \(t\), \(wordlist\) and LLM parameters. Party \(P_2\) holds \(embeddingList\) and embedded parameters. Public insertion ratio $r= 12\%$ as \cite{chang2024postmarkrobustblackboxwatermark}.
\Ensure \(P_1\) receives watermarked text \(t'\)
\State $n \leftarrow \text{CountWords}(t)$  
\State $ k \leftarrow \text{NbWatermarkWords}(\lfloor r \cdot n \rfloor$) 
\State  $k' \leftarrow \text{NbInsertedWords}(3 \cdot k)$ as~\cite{chang2024postmarkrobustblackboxwatermark}.

\State \textit{// Creation of the \texttt{SECTABLE}}
\State $\langle\texttt{SECTABLE}\rangle \leftarrow \Pi_{\text{CSECT}}(wordList, embeddingList)$

\State \textit{// Embed plaintext \(t\) with \texttt{EMBEDDER} }
\State $\langle e_t\rangle \leftarrow \Pi_{\text{Embed}}(t,\ \text{EMBEDDER.params})$

\State \textit{// Words Selection}
\State $\langle \text{sims}\rangle \leftarrow \Pi_{\text{CosineSim}}(\langle e_t \rangle,\ \langle \texttt{SECTABLE} \rangle)$
\State $\langle W_{\text{candidates}}\rangle \leftarrow \Pi_{\text{TopK}}( \langle \text{sims} \rangle,  k')$

\State $\langle \text{simsFiltered}\rangle \leftarrow \Pi_{\text{CosineSim}}( \langle e_t \rangle, \langle W_{\text{candidates}} \rangle)$

\State $\langle W_{\text{filtered}}\rangle \leftarrow \Pi_{\text{TopK}}( \langle \text{simsFiltered}\rangle ,  k )$. \textit{// \texttt{Use cosine similarity as~\cite{chang2024postmarkrobustblackboxwatermark}.}}

\State \textit{// \texttt{INSERTER}}
\State $P \leftarrow$ ``Insert words: $W_{\text{filtered}}$ into the text: $t$"
\State $\langle t'\rangle \leftarrow \Pi_{\text{Insert}}(P, t)$

\State $L_{wm} \gets \text{Save}(W_{\text{filtered}})$ \Comment{Save watermark words}

\State \Return Reveal $t'$
\end{algorithmic}
\end{algorithm}
\end{savenotes}

\begin{savenotes}
\begin{algorithm}[H]
\caption{Protocol $\Pi_{\text{CSECT}}$ Creation of \texttt{SECTABLE}}
\label{alg:sectable}
\begin{algorithmic}[1]
\Require Party \(P_1\) holds the vocabulary $V$. Party \(P_2\) holds the set of embeddings $D$.
\Ensure Secure embedding table $\langle\texttt{SECTABLE}\rangle$, and host mappings $\text{word2idx},\ \text{idx2word}$.

\State \textit{// On Host (e.g., Party \(P_1\)'s machine)}
\State $\text{word2idx}, \text{idx2word} \leftarrow \text{CreateMappings}(V)$ \Comment{Create local word-to-index mappings}
\State $M \leftarrow |V|$ \Comment{Get vocabulary size}

\State \textit{// The following operations are performed in 3PC}
\State $N \leftarrow |D|$ \Comment{Get the total number of embeddings from \(P_2\)}
\State $\text{seed} \leftarrow \text{randomSeed}$ and $\text{key} \leftarrow \text{PRNGKey}(seed)$.
\State $\text{perm} \leftarrow \text{RandomPermutation}(\text{key}, N)$ \Comment{Generate a secure permutation of indices $[0, \dots, N-1]$}
\State $\text{chosen\_indices} \leftarrow \text{perm}[:M]$ \Comment{Select the first $M$ indices.}
\State $\langle \texttt{SECTABLE} \rangle \leftarrow D[\text{chosen\_indices}]$ \Comment{Create the secure table by selecting the corresponding embeddings}

\State \Return $\langle \texttt{SECTABLE} \rangle,\ \text{word2idx},\ \text{idx2word}$
\end{algorithmic}
\end{algorithm}
\end{savenotes}

\subsection{Watermark Detection}
Following PostMark~\cite{chang2024postmarkrobustblackboxwatermark}, we need to compute the presence score in watermark detection:
\begin{equation}
    p = \frac{|\{w \in {\bf w} \text{ s.t. } \exists w' \in t', \text{sim}(w', w) \geq \theta_{sim}\}|}{|{\bf w}|},
\end{equation}
where ${\bf w}$ is the list of watermark words.

\begin{algorithm}[H]
\caption{Protocol $\Pi_{\rm Det}$ Watermark Detection}
\label{alg:detection}
\begin{algorithmic}[1]
\Require Party $P_1$ holds candidate text $t'$ and watermark word list $L_{wm}$, party $P_2$ holds embedded parameters. Similarity threshold $\theta_{sim}=0.85$ and detection threshold $\theta_{det}=45\%$ are constants following~\cite{chang2024postmarkrobustblackboxwatermark}.
\Ensure Boolean indicating if a watermark is detected

    \State Let $W_{cand} \leftarrow \text{extract words from } t'$
    \State $\langle E_{cand}\rangle \leftarrow \Pi_{\text{Embed}}(W_{cand},\ \text{EMBEDDER.params})$
    \State $\langle E_{wm}\rangle \leftarrow \Pi_{\text{Embed}}(L_{wm},\ \text{EMBEDDER.params})$
    
    \State $\langle c\rangle \leftarrow 0$
    \For{each word embedding $e_w \in E_{wm}$}
        \For{each word embedding $e'_{w} \in E_{cand}$}
        \State $\langle c\rangle = \langle c\rangle + (\langle \Pi_{\text{CosineSim}}( e_w, e'_{w}) \rangle \geq \theta_{sim})$
        \EndFor
    \EndFor
    
    \State $\langle p\rangle = \langle c\rangle / |L_{wm}|$. \Comment{Calculate the presence score}
    
    \State \textbf{return} $\langle b\rangle = \langle p\rangle > \theta_{det}$
\end{algorithmic}
\end{algorithm}

\section{Experiments}\label{sec:experiment}

\subsection{Experimental Setup}

\noindent\textbf{Implementation.} We implement \tNAME{} on top of SecretFlow-SPU~\cite{spu}. Our experiments are run on a server with an AMD EPYC 7443P 24-Core Processor with 48 CPUs and 256GB RAM, running Ubuntu 22.04 LTS (Linux 5.15.0-82-generic). We utilized a CPU-only implementation, as the GPU-compatible version of SPU is currently experimental. Our evaluations are on GPT-2 base: (1) \tNAME{} versus a plaintext CPU baseline; (2) three networks with bandwidth and latency: localhost = ($26$Gbps, $0.05$ms), LAN = ($1.5$Gbps, $1.5$ms), and WAN = ($400$Mbps, $10$ms).

\noindent \textbf{Models \& Datasets.}
We use \texttt{intfloat/e5-base}~\cite{wang2022text} as our embedding model and variants of GPT-2~\cite{radford2019language} as the \texttt{INSERTER}. Since we use a different embedding model from PostMark~\cite{chang2024postmarkrobustblackboxwatermark}, we construct our own vocabulary embedding set $D$ from the WikiText-103 corpus~\cite{merity2016pointersentinelmixturemodels}.

\subsection{Performance of Secure Watermarking}
We evaluate the performance of \tNAME{}'s insertion and detection phases in respective \S~\ref{sec:exp-insertion} and \S~\ref{sec:exp-detect}.

\subsubsection{Performance of Secure Insertion}\label{sec:exp-insertion}
We evaluated the GPT-2 base model with inputting 50 tokens under Localhost, LAN, and WAN conditions to benchmark the performance of pipeline operations: \texttt{Embed}, \texttt{Cosine}, \texttt{Topk}, and \texttt{Insert} in Table~\ref{tab:network_impact}. 
Firstly, $\Pi_{\rm MarkIn}$ invokes over one magnitude more overhead than plaintext on CPU.
Compared to the local setup, we require a modest overhead of approximately $1.6\times$ for the \texttt{Embed} and $1.2\times$ \texttt{Insert} in LAN. Moreover, in a WAN setting, high latency dramatically increases the cost, with the communication-heavy \texttt{Embed} becoming nearly $16.5\times$ slower than the local \tNAME{}. For \texttt{Cosine} and \texttt{Topk}, it is $35\times$ and $123\times$ slower. This demonstrates that network latency is a critical factor that magnifies the overhead of secure protocols.

\begin{table}[ht]
\centering
\caption{Performance of Protocol $\Pi_{\rm MarkIn}$ of \tNAME{} with GPT-2 base across different networks. All times are in seconds. The \texttt{Insert} in WAN was not completed in 1 hour.}
\vspace{-5pt}
\label{tab:network_impact}
\resizebox{\linewidth}{!}{
\begin{tabular}{lccccc}
\toprule
\multirow{2}{*}{\textbf{Operation}} & \multirow{2}{*}{\textbf{CPU (s)}} & \multicolumn{4}{c}{$\Pi_{\rm MarkIn}$} \\ \cline{3-6}
& & \textbf{Lo  (s)} & \textbf{LAN (s)} & \textbf{WAN (s)} & \textbf{Comm. (MB)}\\
\midrule
\texttt{Embed}  &1.191    & 23.76 & 37.95 & 393.4 & 2285 \\
\texttt{Cosine} &0.038   & 0.047 & 0.060 & 1.775 &  $0.221$ \\
\texttt{Topk}   &0.006  & 0.008 & 0.012 & 0.986 & 0.429 \\
\texttt{Insert} &57.704   & 1267 & 1551 & $>1$ hour & $71431$ \\
\bottomrule
\end{tabular}}
\end{table}

We also measure the overhead of \tNAME{} in Table~\ref{tab:network_impact}.
Similarly to time profiling, the Insert operation accounts for more than $90\%$ of the communication overhead.

\subsubsection{Performance of Secure Detection}\label{sec:exp-detect}
We benchmark the performance of $\Pi_{\rm Det}$ in Table~\ref{tab:detection_overhead}.
Indeed, $\Pi_{\rm Det}$ is substantially slower than the plaintext baseline (which only requires $1.17$ seconds), incurring approximately $145\times$ overhead.
This significant slowdown is consistent with~\S~\ref{sec:exp-insertion}, as detection relies heavily on embedding and similarity calculations, which are expensive within MPC.
Also, network latency has a substantial effect on the running time of \tNAME{} detection.

\begin{table}[!t]
\centering
\caption{Performance of Protocol $\Pi_{\rm Det}$ of \tNAME{} under different networks.}
\vspace{-5pt}
\label{tab:detection_overhead}
\resizebox{\linewidth}{!}{
\begin{tabular}{lccccc}
\toprule
\textbf{Operation} & \textbf{CPU (s)} & \textbf{Lo  (s)} & \textbf{LAN (s)} & \textbf{WAN (s)} & \textbf{Comm. (MB)}\\
\midrule
$\Pi_{\rm Det}$  & $1.17$   & $172$ & $321.6$ & 4037 &  26795  \\
\bottomrule
\end{tabular}}
\vspace{-10pt}
\end{table}

\subsection{Evaluation of Effectiveness and Robustness} \label{subsec:robustness}

\noindent{\bf Example of Insertion Outputs.}
To empirically show that our \tNAME{} is numerically precise in the watermarking insertion, we first give a representative example to compare our generated outputs to the plaintext CPU baseline.
As illustrated in Figure~\ref{fig:example}, \tNAME{} produces semantically identical outputs, confirming that \tNAME{} correctly executes the watermarking logic while preserving privacy.
More examples can be reproduced using our open-source repository.

\begin{figure}[!t]
    \centering    
    \begin{tikzpicture}
        \node[rectangle, rounded corners, fill=white, text width=\linewidth, inner sep=0pt, align=left] (box) {
        \begin{minted}[
            linenos,               % Show line numbers
            framesep=2mm,          % Frame separation from the code
            numbersep=4pt,         % Line number separation
            bgcolor=verylightgray, % Light background color
            breaklines,            % Enable line breaking
            autogobble,            % Automatically trim leading spaces
            fontsize=\scriptsize % Set font size to footnotesize
        ]{python}
[Candidate Words]
Top 6: ['lawyer', 'decade', 'lyrical', 'room', 'print', 'stance']
Filtered Top 2: ['lyrical', 'lawyer']

[PrivMark / Plaintext Insertion Output]
'Insert the following words: [lyrical', 'lawyer'] into the text:
"We are not! Like men in the story of the Good Samaritan, they pass by on the other side..."'
The first sentence of the sentence is a reference to the fact that the Good Samaritan is a man.  The second sentence is a reference to the fact that the Good Samaritan is a woman.
\end{minted}
};
\end{tikzpicture}
\vspace{-20pt}
\caption{Example of insertion outputs of \tNAME{}. Plaintext baseline produces the same output.}
    \label{fig:example}
    \vspace{-10pt}
\end{figure}

\noindent{\bf Experimental Evaluation.}
Finally, we evaluate our end-to-end effectiveness and robustness by measuring the watermark's detection rates.
We generate around 70 examples with Google Gemini 2.5 Pro \cite{comanici2025gemini25pushingfrontier}. Examples counts examples from different domains (i.e., technology, history, astronomy, literature, etc). 
The detailed examples are illustrated in our repository.
The measure methodology involved a true positive test to measure robustness against attacks and a false positive test to measure accuracy: 
\romannumeral1) For the true positive test, we check for the watermark in the original text, a paraphrased version, and a version where watermark words were partially removed. 
\romannumeral2) For the false positive test, we check for a specific set of watermark words in unrelated texts (from a different domain than the original watermark wordlist text).

As show in Table~\ref{tab:detection_rates}, the watermark is perfectly detected (100\%) in the original text. It also shows high robustness to paraphrase attacks (91.2\%) and moderate robustness to removal attacks (79.4\%). The false positive rate on unrelated texts is 13.2\%.
Compared to plaintext, we achieve identical metrics. This validates our effectiveness and robustness guarantees with privacy-preservation.

\begin{table}[!t]
\centering
\caption{Detection rates under different attacks.}
\vspace{-5pt}
\label{tab:detection_rates}
\begin{tabular}{lcc}
\toprule
\textbf{Test Scenarios} & Plaintext & \tNAME{} \\
\midrule
Original (True Positive) & 100.0\% & 100.0\% \\
Paraphrase Attack & 91.2\% & 91.2\% \\
Removal Attack & 79.4\% & 79.4\% \\
\midrule
Unrelated Text (False Positive) &  13.2\% & 13.2\% \\
\bottomrule
\end{tabular}
\vspace{-15pt}
\end{table}

\section{Conclusion}\label{sec:conclusion}

We propose \tNAME{}, the first framework that integrates secure multi-party computation (MPC) with large language model (LLM) watermarking. Our work addresses the critical need for content traceability and ownership verification in privacy-sensitive scenarios where models, data, and watermarking keys cannot be exposed. 
This work serves as a foundational step, demonstrating the feasibility of MPC-based LLM watermarking. 
Future work can focus on optimizing the cryptographic protocols and efficient implementations to make private, robust, and scalable watermarking for modern LLMs.

\bibliographystyle{IEEETran}
\bibliography{IEEEabrv,ref}

\appendix

\section{Survey of black-box LLM Watermarking Techniques}\label{sec:wm_solution}

In this appendix we compare representative text‐watermarking methods along the criteria of fidelity, robustness, efficiency, and undetectability. Table~\ref{tab:llm_watermarking_part1} summarizes key metrics from each technique. Below we briefly describe each approach and its experimental performance.

\subsection{Unicode Watermarking~\cite{sato2023embarrassinglysimpletextwatermarks}}
Unicode‐based watermarks (e.g., zero‐width or homoglyph insertions) leave the visible text unchanged, so fidelity is perfect (BLEU and PPL drop 0\%). These marks are extremely fragile: standard text normalization (e.g., stripping formatting or normalizing whitespace) removes them entirely, so robustness is ``very weak''. Insertion and detection are trivial (string operations only), giving extremely high efficiency. The watermark is imperceptible to human readers (appearance unchanged), though detectable by a specialized validator with near 100\% true positive on clean text (and essentially 0\% false positives).

\subsection{Yang \emph{et al.} \cite{yang2023watermarkingtextgeneratedblackbox}}
Yang \emph{et al.} propose a black‐box scheme that encodes each word as a random bit and substitutes ``0''‐bit words with context‐appropriate synonyms (via BERT/WordNet). This preserves meaning very well, so fidelity is high (semantics are largely unchanged). Robustness is moderate: the watermark is hard to erase without altering semantics, but paraphrasing attacks can largely defeat it (e.g., detection rate, F1-score, drops from $\sim$81\% on clean text to $\sim$30\% after paraphrase of English texts - Figure 10-11). Efficiency is moderate: encoding requires synonym lookup with a BERT model, which is more expensive than string‐level methods but much cheaper than invoking an LLM. Undetectability is high: humans see only plausible synonyms, so the watermark is essentially imperceptible to readers.

\subsection{DeepTextMark \cite{munyer2024deeptextmarkdeeplearningdriventext}}
DeepTextMark uses Word2Vec and sentence embeddings to apply many small synonym swaps, and trains a Transformer to detect the pattern. It achieves very high fidelity (imperceptible edits): reported semantic similarity (mSMS) is $\sim$0.99, i.e., almost no meaning shift. Detection accuracy is also very high: in experiments, sentence‐level detection accuracy approaches $\sim$95\% or more (near 100\% with multiple sentences). The watermark survives moderate editing: e.g., removing or adding a few sentences only slightly degrades performance (the AUC remains high, see Table~3–5). Efficiency is good: insertion takes on the order of 0.28\,s per sentence (single‐core CPU) and detection $\sim$0.002\,s, making it practical. In summary, DeepTextMark achieves high fidelity and robustness with reasonably low overhead. The watermark is imperceptible to humans (word substitutions are context‐appropriate).

\subsection{PostMark \cite{chang2024postmarkrobustblackboxwatermark}}
PostMark is a black‐box method that selects an input‐dependent set of ``watermark words'' via semantic embeddings and uses an instruction‐tuned LLM to rewrite the text to include those words. It produces an imperceptible watermark (cosine similarity of embeddings original vs.\ watermarked is $\sim$0.94--0.95). Detection is extremely strong: at 1\% FPR, PostMark@12 achieves $\sim$99--100\% TPR on clean text, and still $\sim$52--59\% TPR after GPT‐3.5 paraphrasing. This greatly outperforms simpler schemes (e.g., Yang’s) which collapse near 0\% after paraphrase. The trade‐off is efficiency: PostMark is slow and costly, as it makes multiple LLM calls. In one report, watermarking a $\sim$280‐token passage took $\sim$36\,s (on Llama‐3‐8B) and cost about \$1.2 per 100 tokens. Detection itself is trivial once watermarked (just check for the chosen words). Undetectability is high: human evaluation found that the inserted words cannot be reliably spotted, and text quality (coherence, relevance) remains high.

\subsection{REMARK-LLM \cite{299615}}
REMARK-LLM is an end‐to‐end neural watermark that learns to embed a binary message via token replacement and decodes it with a neural extractor. It is explicitly trained to preserve meaning. Experiments show REMARK-LLM can embed, e.g., 64 bits in 640 tokens with only slight impact on semantics. Semantic fidelity is very high (BERTScore $\approx$0.92--0.94 vs.\ original). Watermark extraction is extremely reliable: most embedded bits are recovered (about 95--98\% accuracy on clean text, i.e., word‐error $\sim$2--5\%). Robustness is also strong: after deletion/addition attacks, detection AUC remains around 0.88--0.90, far above prior neural methods. Efficiency is moderate: REMARK uses a trained model (requires GPU $\sim$5.8\,GB) but no LLM calls. On a T5‐large backbone it takes about 1.2\,s to watermark an 80‐token segment. Overall, REMARK-LLM offers high payload, high fidelity, and good robustness.

\subsection{Generation of examples}\label{sec:examples}
We generate example thanks to Google Gemini 2.5 Pro. Examples are as follows.

\begin{figure*}[!t]
    \centering    
    \begin{tikzpicture}
        \node[rectangle, rounded corners, fill=white, text width=\linewidth, inner sep=0pt, align=left] (box) {
        \begin{minted}[
            linenos,               % Show line numbers
            framesep=2mm,          % Frame separation from the code
            numbersep=4pt,         % Line number separation
            bgcolor=verylightgray, % Light background color
            breaklines,            % Enable line breaking
            autogobble,            % Automatically trim leading spaces
            fontsize=\tiny % Set font size to footnotesize
        ]{python}
    "id": 1,
    "word_count": 105,
    "watermark_count": 12,
    "watermark_words": ["monumental", "classical", "qubits", "superposition", "simultaneously", "entanglement", "incredible", "materials", "encryption", "maintaining", "enormous", "decoherence"
    ],
    "candidate_text": "Quantum computing represents a monumental leap forward from classical computing. Instead of bits, which store information as either 0 or 1, quantum computers use qubits. These qubits can exist in a superposition, a combination of both states simultaneously. This fundamental property, along with entanglement, allows quantum machines to process a vast number of calculations in parallel. The potential applications are incredible, ranging from drug discovery and materials science to breaking current encryption standards. However, building and maintaining stable quantum computers presents enormous technical challenges, demanding extremely low temperatures and isolation from environmental interference to prevent decoherence.",
    "removing_attack": "Quantum computing represents a leap forward from computing. Instead of bits, which store information as either 0 or 1, quantum computers use qubits. This property allows quantum machines to process a vast number of calculations in parallel. The potential applications range from drug discovery and science to breaking current standards. However, building stable quantum computers presents technical challenges, demanding extremely low temperatures and isolation from environmental interference to prevent issues.",
    "paraphrase_attack": "A massive advancement over traditional computing is offered by quantum computation. Quantum computers utilize quantum bits rather than standard bits that hold data as a 0 or 1. These can be in a mixed state, representing both values at the same time. This core characteristic, combined with quantum linking, enables these devices to perform a huge volume of computations concurrently. The possible uses are amazing, spanning from creating new medicines and substances to defeating modern security codes. Nevertheless, constructing and sustaining stable quantum systems poses immense engineering difficulties, requiring frigid conditions and shielding from outside disturbances to avoid losing the quantum state.",
    "domain": "Technology"
  },
  {
    "id": 2,
    "word_count": 102,
    "watermark_count": 12,
    "watermark_words": ["singular", "complex", "incessant", "instability", "crippled", "innovation", "indefensible", "expenditure", "decimated", "gradually", "foundations", "fragmentation"
    ],
    "candidate_text": "The decline of the Western Roman Empire was not a singular event but a complex, centuries-long process. Historians attribute the collapse to a combination of factors, including incessant barbarian invasions, internal political instability, and severe economic troubles. The over-reliance on slave labor crippled technological innovation, while vast, indefensible borders required an enormous military expenditure that the state could not sustain. Furthermore, a series of plagues decimated the population, reducing the tax base and the number of available soldiers. This perfect storm of challenges gradually eroded the empire's foundations, culminating in its eventual fragmentation.",
    "removing_attack": "The decline of the Western Roman Empire was not an event but a long process. Historians attribute the collapse to a combination of factors, including barbarian invasions, internal political issues, and severe economic troubles. The over-reliance on slave labor hindered progress, while vast borders required a military that the state could not sustain. Furthermore, a series of plagues reduced the population, the tax base and the number of available soldiers. This perfect storm of challenges eroded the empire's base, culminating in its eventual breakup.",
    "paraphrase_attack": "The fall of the Western Roman Empire was a multifaceted, protracted procedure rather than a single occurrence. Scholars suggest that a mix of elements led to the downfall, such as constant barbarian incursions, domestic political turmoil, and grave financial difficulties. A heavy dependence on enslaved workers stifled technological advancement, and extensive, unprotectable frontiers demanded a massive military budget the government could no longer afford. Additionally, a succession of epidemics wiped out many people, shrinking tax revenue and potential recruits. This confluence of crises slowly weakened the empire's core, leading to its ultimate dissolution.",
    "domain": "History"
  }
\end{minted}
};
\end{tikzpicture}
\vspace{-15pt}
\caption{Example of insertion outputs of \tNAME{}.}
    \label{fig:querytrans}
    \vspace{-5pt}
\end{figure*}

\end{document}